\documentclass[twocolumn,showpacs,preprintnumbers,amsmath,amssymb,superscriptaddress]{revtex4-1}

\usepackage{graphicx}
\usepackage{dcolumn}
\usepackage{bm}
\usepackage{textcomp}
\usepackage{color}
\usepackage{xcolor}
\usepackage{ulem}
\usepackage{amsmath,amsfonts,amssymb}

\begin{document}
\preprint{Submitted to Phys. Rev. B}


\title{Understanding XANES spectra of two-temperature warm dense copper \\
 using \textit{ab initio} simulation.}

\def\celia{Univ. Bordeaux, CNRS, CEA, CELIA, UMR 5107, F-33400 Talence, France}
\def\cea{CEA-DAM-DIF, F-91297 Arpajon, France}
\def\loa{LOA, ENSTA ParisTech, CNRS, EcolePolytechnique, Universit\'e Paris-Saclay, 91120 Palaiseau, France}

\author{N. Jourdain}
\affiliation{\cea}
\affiliation{\celia}
\author{V. Recoules}
\affiliation{\cea}
\email{vanina.recoules@cea.fr}
\author{L. Lecherbourg}
\affiliation{\cea}
\affiliation{\loa}
\author{P. Renaudin}
\affiliation{\cea}

\author{F. Dorchies}

\affiliation{\celia}

\date{\today}

\begin{abstract} 
Using \textit{ab initio} molecular--dynamics simulations combined with linear--response theory, we 
studied the x--ray  absorption near--edge spectra (XANES) of a two-temperature dense
copper plasma. As the temperature increases, XANES spectra exhibit a pre--edge structure balanced by
a reduction of the absorption just behind the edge.  
By performing systematic simulations for various thermodynamic
conditions, we establish a formulation to deduce the electronic temperature $T_e$ directly from the spectral integral of
 the pre--edge that can be used
for various thermodynamic conditions encountered in a femtosecond heating experiment where
 thermal non--equilibrium and expanded states have to be considered.
\end{abstract}
\maketitle

Time resolved X-ray absorption near-edge spectroscopy (TR-XANES) was recently extended to the Warm
Dense Matter (WDM) regime \cite{Dorchies2016}. This highly transient physics has required specific
development to get an appropriate time-resolution (picosecond or less) on XANES measurements.
TR-XANES gives a complete picture by probing simultaneously the valence electrons and the atomic
local arrangement modifications in WDM situations well beyond the melting. However, the physical
interpretation of XANES spectra is not straightforward and a strong connection between theory and
experiment is needed to extract information. This issue has been addressed using \textit{ab initio} molecular
dynamics (AIMD) simulations, providing a consistent description of both the electronic and ionic structures,
together with X-ray absorption spectra calculation \cite{Recoules_2009, Cho_2011}. 

In the past decade, both theoretical and experimental XANES spectroscopy was used to get a deep
understanding of the electronic structure modification of metals in the WDM.  On aluminum,  XANES
modification during the solid--liquid--vapor phase transition was observed with picosecond time-resolution \cite{Dorchies_2011}.
With the support of AIMD simulations, this has shown the connection between
electronic and ionic modifications. Comparison between measurements and
calculations for molybdenum have demonstrated that XANES spectra can be simply interpreted in terms of
electron DOS modification when solid molybdenum turns to WDM \cite{Dorchies_PRB2015}. Besides
the understanding of the phenomena involved, it is possible to have access to the timescales of phase
transitions \cite{Mancic_2010, Dorchies_2011, Leguay_2013}. XANES spectra were also used to shed light on the DOS
modification induced by laser shock compression in metals such as aluminum \cite{Benuzzi_2011,
Levy_2012} and iron \cite{Marini_2014, Harmand_2015}.

When WDM is transiently produced by femtosecond laser heating, matter faces non equilibrium situations. It is a great scientific challenge to independently resolve the electron and ion dynamics. 
In principle, TR--XANES can address this physics, but it is necessary to disentangle the corresponding features in XANES spectra,
which depend on the considered element.
In some light metals such as Be \cite{Schwanda_1992} and Al \cite{Dorchies_PRBb2015}, the
electronic temperature can be retrieved directly from the slope of the absorption $K$ edge.
In most metals with a localized $d$ band,  the situation is more complex and this methodology cannot
be applied. Recent experiments have been dedicated to femtosecond laser heated warm dense copper on a synchrotron beamline \cite{Cho_2011, Cho_2015}. 
Using 2 ps time resolved X-ray spectroscopy, Cho et al investigated the modification of XANES spectra near $L_{2,3}$ edge.
By direct comparison between measured spectra and spectra computed using AIMD, they retrieved the time evolution of the electron temperature and compare it to the behaviour obtained using a two-temperature models. 

The purpose of the present paper is to go further in the calculations and analysis of warm dense copper using AIMD simulations.
The interpretation of XANES spectra especially for transition metal is non trivial and require a careful analysis.
The relation between the electronic temperature $T_e$ and pre--edge peak is reproduced. A quasi linear function is extracted to deduce absolute values of $T_e$ (up to 3 eV) from the spectral integration of the pre--edge. The impacts of ion temperature $T_i$ and density are carefully and independently studied. This shows the validity of this function in the 
various thermodynamic conditions encountered in a femtosecond heating experiment, including strong thermal non
equilibrium. This provides a practical $T_e$ diagnostic to analyze any XANES experiment without the need of additional AIMD calculations.
It has been used in a recently published paper that revisits the electron--ion thermal equilibration dynamics on a table--top set up, 
emphasizing the critical role of target expansion \cite{XANES_Cu}. Other features are identified above L-edge, in close relation with the crystalline structure.

\section{XANES spectrum at ambient condition} 
Before analyzing the modification of XANES spectra induced by electron temperature ($T_e$), ion temperature ($T_i$), and density $_rho$, we first
need a clear understanding of the connection between XANES spectra and the electronic structure of
copper at ambient condition. This is also a way to check the validity and the precision of the method we
use.

\subsection{Computational method} 
To obtain X-ray absorption spectra of warm dense copper, we first need to generate a set of atomic
configurations for a given ($\rho$,$T_i$,$T_e$) condition.  AIMD simulations are performed using the \textit{ab initio}
plane wave Density Functional Theory (DFT)  code \textsc{Abinit}~\cite{Abinit, Gonze_2009, Bottin_2008}.
DFT is applied together with Local Density Approximation \cite{Perdew_1992}.  For non-equilibrium AIMD
simulations the ion and electron temperatures are controlled independently. While the electron temperature
is fixed for every time step in the simulation, the ion temperature is controlled in the iso-kinetics ensemble
where the velocity is rescaled at every time step to maintain the desired temperature.

We start the simulations with 108 atoms initially arranged in a perfect fcc lattice. The equilibration for a
given temperature is then monitored by looking at the pressure. Before reaching a stable ionic structure,
the system relaxes for a few hundreds of time steps during which the pressure increases. After equilibration,
it fluctuates around a well--defined value. At the end, the initial configuration is propagated up to $2\;\rm{ps}$
using time steps of  $2\;\rm{fs}$ for the low temperature and $0.5\;\rm{fs}$ for temperatures above 1 eV.
We consider electronic states occupied down to $10^{-6}$ electrons per states which means around 1 600 bands are explicitly
computed for the highest temperature considered here.  All molecular dynamics calculations were performed
at the $\Gamma$ point to represent the Brillouin zone. The $\Gamma$ point sampling is expected to be a
good approximation for the calculation of the structure and dynamics of copper. We use for this element a
PAW data set generated using 11 outer electrons ($3d^{10}4s^1$) and  a cut--off radius of $2\;\rm{bohr}$
that has been benchmarked against physical properties obtained from experiments \cite{Dewaele_2008}.
The plane wave cut--off is $15\;\rm{Ha}$.

We then select equally spaced ionic configurations along the equilibrated part of the trajectory and compute
for each of them the corresponding electronic structure with a DFT calculation. These \textit{ab initio} electronic
structure calculations are performed on a $3 \times 3 \times 3$ Monkhorst-Pack \textbf{k}--point grid and
with 2 500 electronic bands in order to obtain eigenstates converged up to $50\;\rm{eV}$ above the
chemical potential. For this part, the PAW data set includes semi--core states with 19 outer electrons
($3s^23p^63d^{10}4s^1$) and a cut--off radius of $2\;\rm{bohr}$ \cite{Dewaele_2008}. Two projectors are
used for $s$ and $p$ channels, and three projectors for the $d$ channel. The plane wave cut--off is
$25\;\rm{Ha}$. These electronic structure calculations provide all the required outputs to compute the
absorption cross section.

The absorption cross section is calculated in a single electron picture : an electron make  the transition 
from a core orbital $\phi_{core}$ to an excited state $\psi_{n,\bf k}$. The transition energy corresponds to the energy difference between these orbitals. 
Then, for a given $\bm{k}$-point, the absorption cross section is expressed as \cite{Taillefumier_2002}:
\begin{eqnarray}
   \sigma_{\bf k}(\omega)=& &4\pi^2\omega\sum_{n}[1-f(\epsilon_{n,\bf k})]\times \nonumber \\
    & &|\langle\psi_{n,\bf k}\vert\vec\nabla\vert\phi_{core}\rangle|^2\delta(\epsilon_{core}-\epsilon_{n}-\hbar\omega)
\label{eq:sigma}
\end{eqnarray}
We employ atomic units, with the electron charge $e$, Planck's constant $\hbar$, the electron mass
$m_e$, and the fine structure constant $\alpha$ all set to unity. The $n$ summation ranges over the
discrete bands (orbitals). $f(\epsilon_{n,\bf k})$ is the Fermi-Dirac occupation factor corresponding to
the energy $\epsilon_{n,\bf k}$ of the $n$th band and for the {\bf k}--point ${\bf k}$. The total cross
section is obtained by direct summation over all necessary {\bf k}--points.

To take into account the natural linewidth of the levels involved in the transition and the resulting
spectral broadening of the absorption edges, we use a Lorentzian function to model the delta function of
equation \ref{eq:sigma}:
\begin{equation}
  \delta(\epsilon_{core}-\epsilon_{n}-\hbar\omega)=\frac{1}{\pi}
  \frac{\Gamma_f(\omega)}{\Gamma_f(\omega)^2+(\epsilon_{core}-\epsilon_{n}-\hbar\omega)^2}
\label{eq:Lorentz}
\end{equation}
This Lorentzian function has an energy-dependent width $\Gamma_f$ according to the reference
\cite{Bunau_2013} which reads : 
\begin{equation}
\Gamma_f(\omega)=\left\{
  \begin{array}{ll}
   0, & \hbar\omega < E_F\\
   \Gamma_{\bm{hole}}+\gamma(\omega), & \hbar\omega > E_F\\ 
  \end{array}
 \right .
  \label{eq:grandGamma}
\end{equation}
This reproduces structures at the $L_2$ edge broader than the ones at the $L_3$ edge.

In equation \ref{eq:grandGamma}, $\Gamma_{\bm{hole}}$ is the core level spectral width. The $\gamma$
function is an arctangent like function that gives a smooth broadening from $\Gamma_{\bm{hole}}$ to a
chosen $\Gamma_{\bm{max}}$ value, and is expressed as:
\begin{equation}
  \gamma(\omega)=\Gamma_{\bm{max}}\left[ \frac{1}{2}+\frac{1}{\pi}\arctan\left(e-\frac{1}{e^2}\right) \right]
  \label{eq:petitgamma}
\end{equation}
where $e=(\omega-E_F)/(E_{ctr}-E_F)$ and $E_{ctr}$ is the inflection point of the arctangent function.

\begin{figure}
  \centering
  \includegraphics[width=0.98\linewidth]{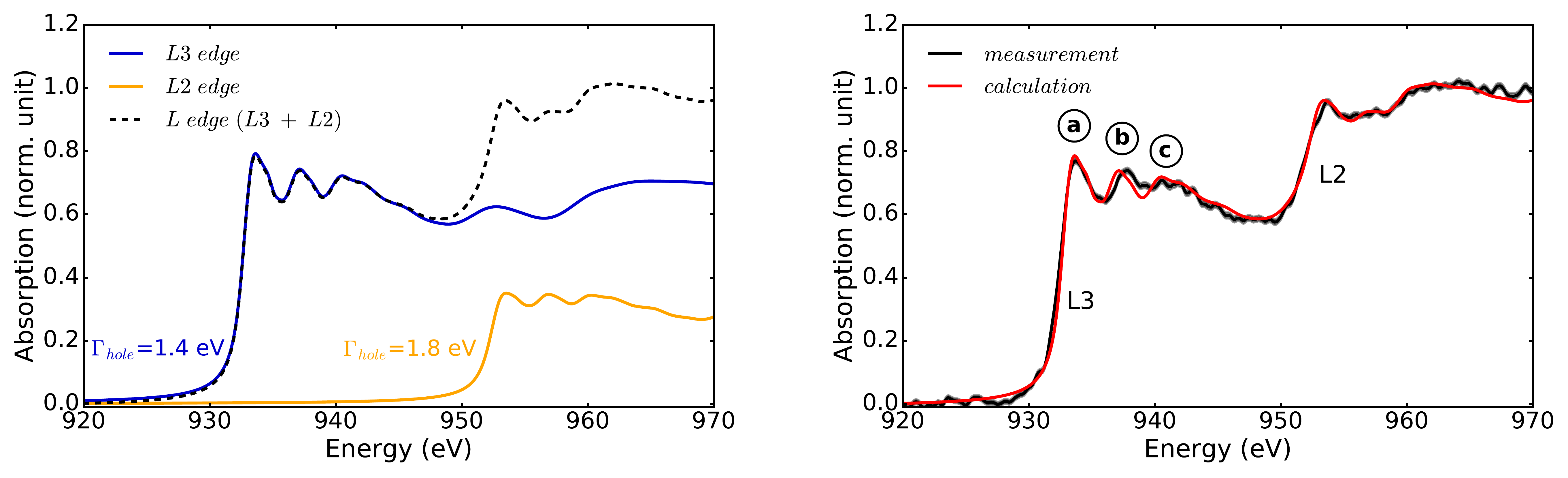}  
  \vspace*{0.5cm}\caption{(color online). \textit{Left panel} : Construction of the total XANES spectrum at $L_{2,3}$ absorption
  edges from the spectrum calculated at $L_3$ edge for the case in ambient conditions ($\rho_0$, $T_e=T_i=$
  300 K). \textit{Right panel} : Calculation (red curve) of the XANES spectrum at $L_{2,3}$
  absorption edges of copper at normal solid density $\rho_0$ = 8.9 g/cm$^3$ and at $T_e=T_i=$ 300 K
  compared to experiment in the same conditions (black curve).}
  \label{Cu_rho0_300K_construct}
\end{figure}

The core states $|\phi_{core}\rangle$ are obtained from an all-electron atomic calculation performed
during the generation of our PAW data set using Atompaw code \cite{Holzwarth_2001}. For the $L_{2,3}$
absorption edge which we consider here, we need the $\phi_{2p}$ orbitals. The radial part of $2p_{1/2}$
and $2p_{3/2}$ core wave function are taken identical.  The spin-orbit splitting of the $2p$ states is
introduced by shifting the calculated absorption spectrum by the value of the energy separation between
the $L_{3}$ and the $L_{2}$ edges. It can be obtained from a DFT relativistic all--electron calculation and is
equal to $20 \ \rm{eV}$. The two edges have similar structures, but their intensities are related by a factor
of 2 (the statistical branching ratio). Left panel of figure \ref{Cu_rho0_300K_construct} shows how the $L_2$ edge is
constructed from the $L_3$ edge for the XANES spectrum at ambient conditions. 

During the experimental
measurements of the XANES spectra, the system is modified by the probe itself as an inner--shell electron
is removed. In many cases \cite{Mauchamp_2009}, it is necessary to include in the simulation cell one atom
described by a PAW data set including a hole in the core shell. This is a way to model the electron--hole
interaction but it neglects many body interactions and dynamic core-hole interaction. 
In the case of copper, the agreement with the experimental spectrum is better without the
consideration of the core hole. This was already observed in ref \cite{Bunau_2013}. 
Quoting Mauchamp \textit{et al.} \cite{Mauchamp_2009}, a core-hole effect is always present in any core-level 
spectroscopy experiment but its effect is not always evidenced and nothing can be said a priori concerning the agreement or not with the experiment.  When building PAW dataset with a hole in the core, there is no mean to test those PAW datasets. The only way is to compare directly the XANES spectra with experiment. This was also the choice in ref \cite{Cho_2011, Cho_2015}.

At the end, we compute absorption cross section for each atom in the simulation box for each snapshot and then do an average on the resulting spectra. 

\subsection{Comparison with experiment}
The right panel of figure \ref{Cu_rho0_300K_construct} shows the comparison between the theoretical and the experimental
\cite{XANES_Cu} XANES spectra in ambient conditions. The absolute position of the edge can not be
formally obtained with the present model. To be compared with the experiment, the theoretical spectrum
for the cold case is shifted to match the experimental energy position of the edge. The same rigid shift
will then be applied to all computed absorption spectra. 

The best agreement with the experimental cold spectrum is obtained using $\Gamma_{\bm{hole}} = 1.4$ eV
and $\Gamma_{\bm{max}} = 5.1$ eV for the $L_3$ edge. For the $L_2$ edge, we use a fixed value of 1.8 eV.
These values of $\Gamma_{hole}$ are close to the one found in the reference \cite{Krause_1979} taken into 
account the spectral resolution of the experimental spectrum. These values are used for all computed 
spectra and can be seen in Figure \ref{Cu_rho0_300K_construct} for both $L_3$ and $L_2$ edges.
 
The agreement between experimental and theoretical spectra is excellent. The first three peaks above the
$L_3$ edge (labeled $(a)$, $(b)$ and $(c)$) and the minimum around 950 eV are well reproduced. The
spectral shape and amplitude of the $L_2$ edge are also well described by the theoretical spectrum.

\begin{figure}
  \centering
  \includegraphics[width=0.8\linewidth]{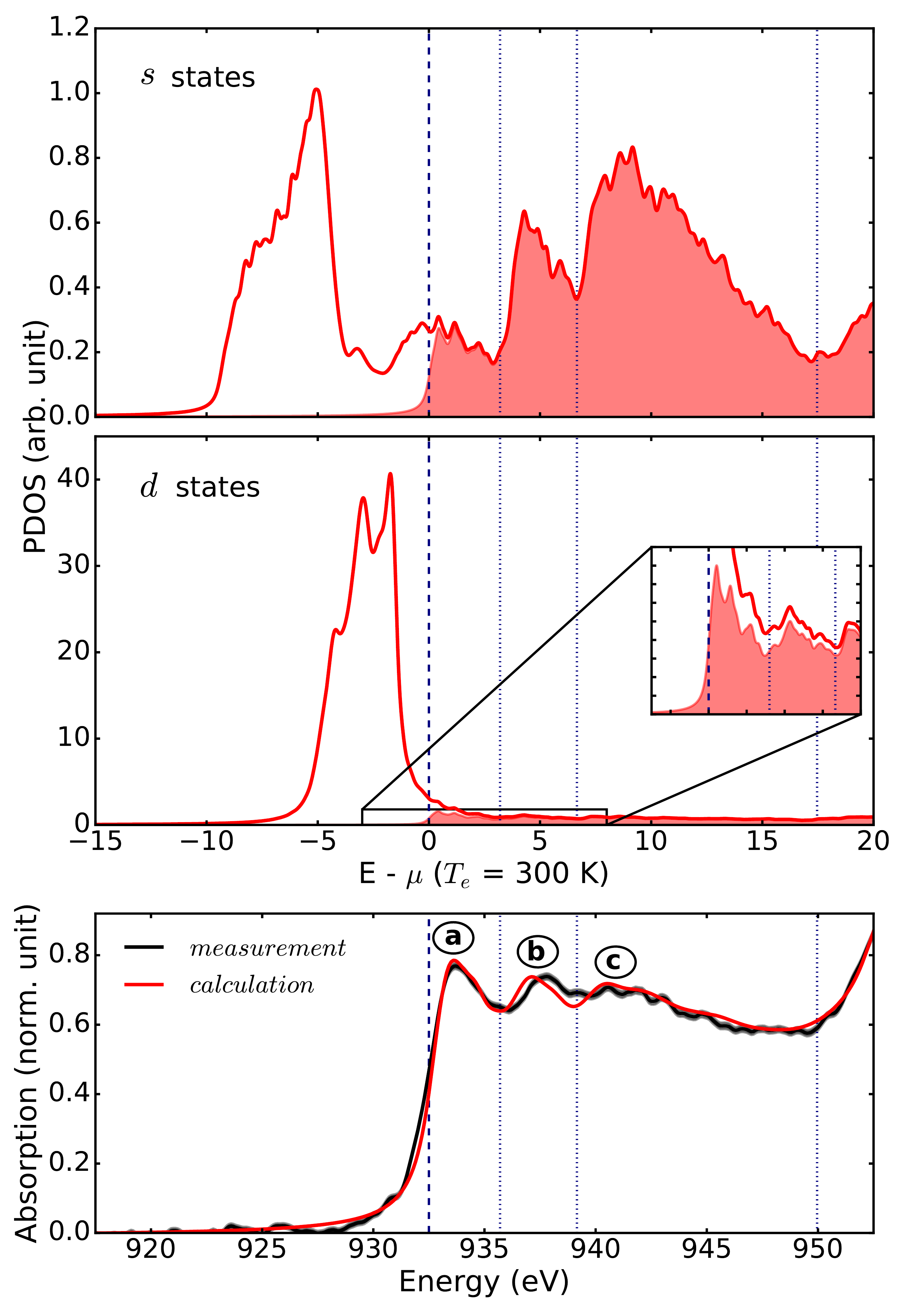}  
  \vspace*{0.5cm}\caption{(color online). \textit{Upper panel: } Projected density of states (PDOS) for
  copper at normal solid density $\rho_0$ and at $T_e=T_i=$ 300 K for $s$ and $d$ channels. The colored
  areas correspond to the vacant electronic states. The $d$ channel is zoomed because of the high amplitude
  of the $3d$-band localized $\sim$ 1.5 eV below $\mu$($T_e=300$ K). \textit{Lower panel: } Calculation of
  the XANES spectrum at $L_{3}$ absorption edges of copper (red curve) compared to experiment in the
  same conditions (black curve).}
  \label{Cu_rho0_300K_pdos}
\end{figure}

$L_{2,3}$ edges probe the unoccupied $s$ and $d$ states. This is illustrated in the Figure
\ref{Cu_rho0_300K_pdos} showing the density of states projected on $s$ and $d$ channels together with
the corresponding XANES spectrum at the $L_3$ edge. $p$ states are not shown here as they do not
contribute to the absorption cross section because of the selection rule $\Delta l = \pm 1$. For copper at
ambient conditions, the $3d$ states are entirely occupied and the XANES spectrum reflects the $s$ and
$d$ type orbitals above the chemical potential $\mu(T_e)$.  The colored areas in Figure
\ref{Cu_rho0_300K_pdos} refer to the unoccupied DOS which is representative of the possible final states
of the photoelectron transition. On this shaded part, we recognize the features in the electron DOS that are
responsible for the three peaks $(a)$, $(b)$ and $(c)$ localized between 930 eV and 940 eV in the XANES
spectrum, as well as the minimum amplitude at $\sim$ 20 eV above $\mu$($T_e=300$ K). 

\section{Shape modifications induced by temperature.}
To identify the shape modifications induced by the rise of the temperature, we carry out calculations of the x-ray
absorption spectrum for several electronic and ionic temperatures at solid density. However, hydrodynamic expansion of thin
targets can become significant after only few picoseconds in ultrafast pump-probe experiments.
To take this into account, we carry out simulations for density below the solid density to see if
this density change has some influence on the modification induced by temperature.

\subsection{Temperature dependence: at thermal equilibrium $\bm{T_e = T_i}$}
\begin{figure}
  \centering
  \includegraphics[width=0.8\linewidth]{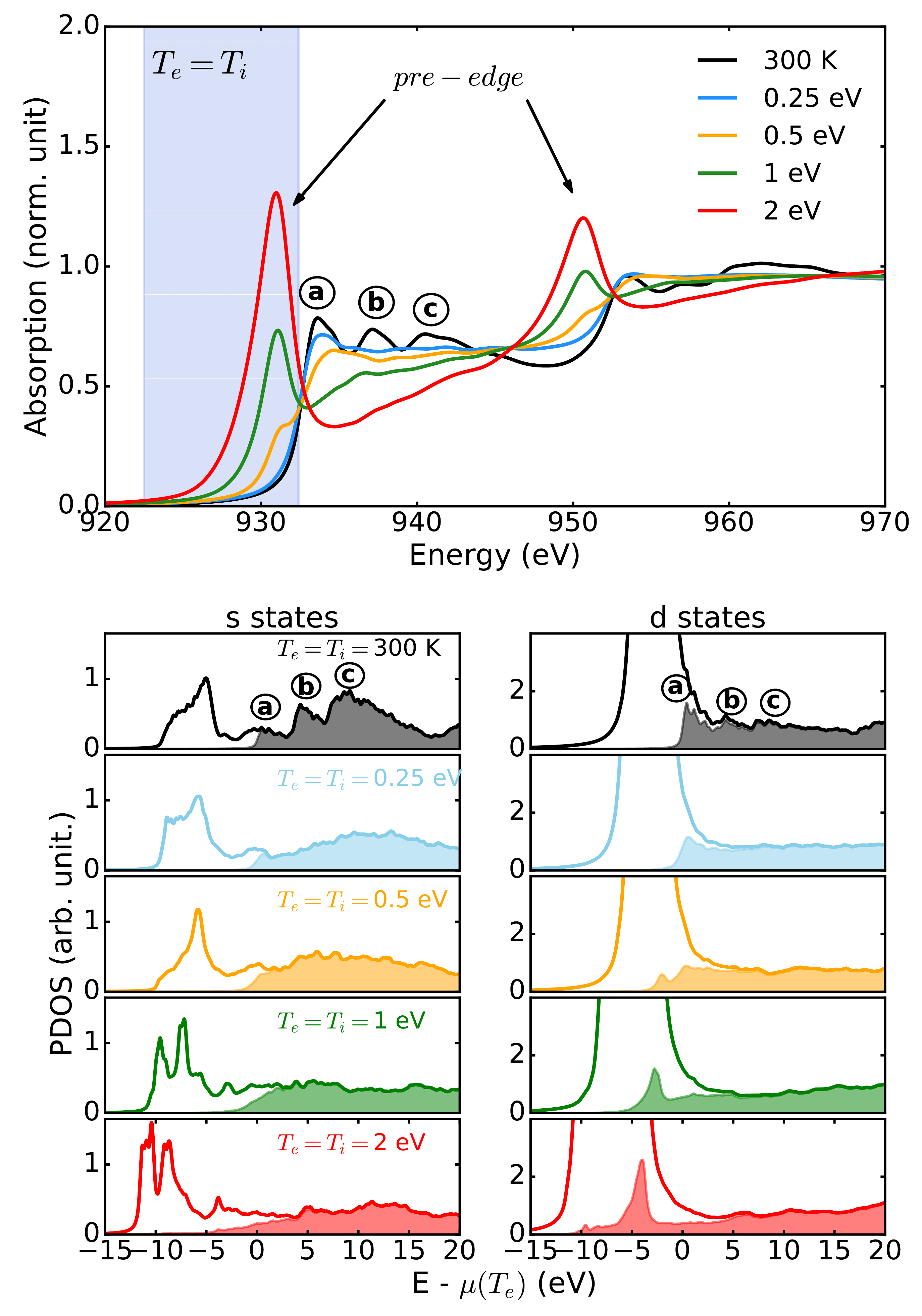}  
  \vspace*{0.5cm}\caption{(color online). \textit{Upper panel: } \textit{ab initio} XANES spectra at $L_{2,3}$ absorption
  edges of copper at normal solid density $\rho_0$ = 8.9 g/cm$^3$ and at thermal equilibrium $T_e = T_i$ up
  to $T_e$ = 2 eV. \textit{Lower panel: } Corresponding projected density of states (PDOS) on $s$ and $d$
  channels. The colored areas correspond to the vacant electronic states.}
  \label{Cu_rho0_X_Te_Ti}
\end{figure}

We start by simulations at thermal equilibrium where $T_e=T_i$ at solid density $\rho_0$ = 8.9 g/cm$^3$.
Computed XANES spectra for temperatures up to 2 eV are reported on the upper panel of Figure
\ref{Cu_rho0_X_Te_Ti}. For the temperature of 0.25 eV, the copper is melted (the melting being expected
at 0.117 eV = 1358 K at ambient pressure).

As the temperature increases, a pre--edge peak appears and is partially balanced by a reduction of the
absorption just behind the edge. This behavior was also observed in ref \cite{Cho_2015} for temperature up
to $2$ eV. The intensity of this pre--edge structure is connected to the value of the electrons temperature
$T_e$. When increasing $T_e$, some electrons from the $3d$-band, fully occupied in ambient conditions, are
promoted towards higher energy states and above $\mu(T_e)$. Thus, the vacant states created below
$\mu(T_e)$ become available for photoionization. $2p\rightarrow 3d$ transitions are then possible, and this 
induces the formation of a pre--edge structure in the XANES spectrum. This was observed that for
transition metals, the $L_{2,3}$ white line intensity can be explained on the basis of $d$ vacancies
\cite{Miyamoto_2010}. This is also true for copper. The size of the pre--edge is connected to the shape of
the electronic occupation, governed by Fermi-Dirac distribution $f(\epsilon,\mu(T_e),T_e)$ defined as:
\begin{equation}
f(\epsilon,\mu(T_e),T_e)=\dfrac{1}{1+ \exp \left(\dfrac{\epsilon-\mu(T_e)}{k_b T_e}\right)}
\end{equation}
It depends on both the energy $\epsilon$, the chemical potential $\mu(T_e)$ and the temperature $T_e$.
The $3d$-band is highly localized at ambient condition and keeps this property even at high temperatures
as can be seen in the lower panel of Figure \ref{Cu_rho0_X_Te_Ti}. Copper has an fcc crystalline structure
at normal conditions, and the local environment does not change strongly upon melting. The empty part of the
$3d$-band consequently appears as a thin peak also beyond melting temperature.

The absorption spectrum becomes however different above the edge. A loss of spectral structures in this
region is clearly observed in the Figure \ref{Cu_rho0_X_Te_Ti} for $T_e=T_i \geq 0.25$ eV, in
close correlation with the loss of DOS structures above $\mu(T_e)$ in the liquid state.  As soon as the
system turns to liquid, the structures labeled $(b)$ and $(c)$ on the XANES spectra and the PDOS vanish.
The first peak $(a)$ stays visible up to 0.5 eV and then disappears as the absorption is reduced in this region.
The structures $(b)$ and $(c)$ might then be used to detect the loss of crystalline structure in copper but 
this would require further analysis to consolidate this observation.
 Note that these (b) and (c) structures are not comparable to EXAFS oscillation observable
around 30eV above the edge which are connected to diffusion of photo--electrons on the neighbour atoms.

\subsection{Electronic temperature dependence at $\bm{T_e \neq T_i=300K}$}
\begin{figure}
  \centering
  \includegraphics[width=0.8\linewidth]{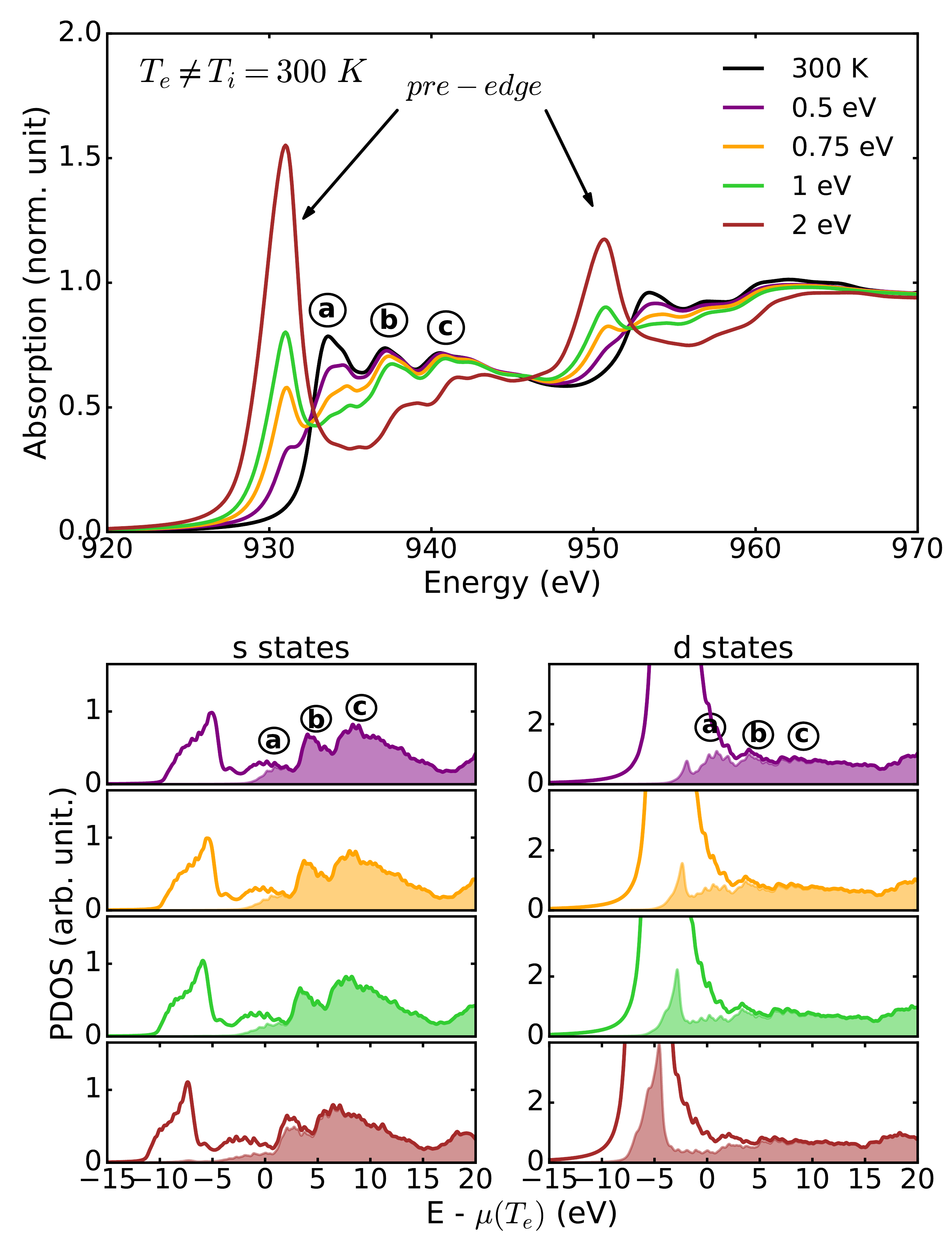}  
  \vspace*{0.5cm}\caption{(color online). \textit{Upper panel: }  \textit{ab initio} XANES spectra at $L_{2,3}$
  absorption edges of copper at solid density $\rho_0$ = 8.9 g/cm$^3$ for strong out-of-equilibrium
  conditions $T_e \gg T_i =$ 300 K up to $T_e$ = 2 eV. \textit{Lower panel: } Corresponding projected
  density of states (PDOS) on $s$ and $d$ channels. The colored areas correspond to the vacant
  electronic states.}
  \label{Cu_rho0_X_Te}
\end{figure}

To enhance the effect of $T_e$ on the spectra without any influence from the atomic structure, we now
focus on the electronic temperature dependence of the XANES spectra by performing simulations at
different $T_e$ up to 2 eV while keeping $T_i$ at 300 K at the solid density $\rho_0$. The ionic structure
is the fcc one obtained for the ions at 300 K. The computed spectra are presented on the upper
panel of Figure \ref{Cu_rho0_X_Te}. As for the case $T_e= T_i$, the broadening of Fermi-Dirac
distribution with $T_e$ always leads to the depopulation of electronic states at the top of the $3d$--band.

The structure referred as $(a)$ on these absorption spectra is strongly reduced as the electronic
temperature $T_e$ increases. The structures labeled $(b)$ and $(c)$ stay clearly visible up to the highest
temperature simulated here.

The lower panel of Figure \ref{Cu_rho0_X_Te} shows the corresponding $s$ and $d$ components of
the PDOS. When increasing the electronic temperature $T_e$ while keeping the same crystalline
atomic structure, the general shape of the $3d$-band is nearly unchanged. This is also true for the
$s$ component. The structures in the projected DOS connected to the peaks $(b)$ and $(c)$ of the
XANES spectra are visible for all temperatures $T_e$. This confirms that the $(b)$ and $(c)$
structures are directly linked to the ionic structure, and that the pre--edge and structure $(a)$ are
representative of the electronic temperature $T_e$.

The DOS shape is different between thermally equilibrated and strong out-of-equilibrium cases
(presented latter in the Fig. \ref{Cu_rho0_eq_vs_HE}).
When copper turns from out-of-equilibrium fcc (ions at 300 K) to liquid at thermal equilibrium
($T_e=T_i$), the $3d$-band stays localized but its shape is modified. That leads to slightly higher
amplitude and thinner pre--edge structures for a given $T_e$ when $T_i$ = 300 K. This can be seen by comparing
the corresponding XANES spectra shown in Figures \ref{Cu_rho0_X_Te_Ti} and \ref{Cu_rho0_X_Te}.
In all cases, areas stay comparable.

\subsection{Temperature dependence for expanded cases $\bm{\rho < \rho_0}$}
The question is now to evaluate the effect of the density on the signature of temperature in XANES
spectra. Simulations for three different density values were therefore performed: at solid density
$\rho_0$ = 8.9 g/cm$^3$, liquid density $\rho_{liquid}$ = 8 g/cm$^3$ and at an expanded density
of 6 g/cm$^3$.

\begin{figure}
  \centering
  \includegraphics[width=0.7\linewidth]{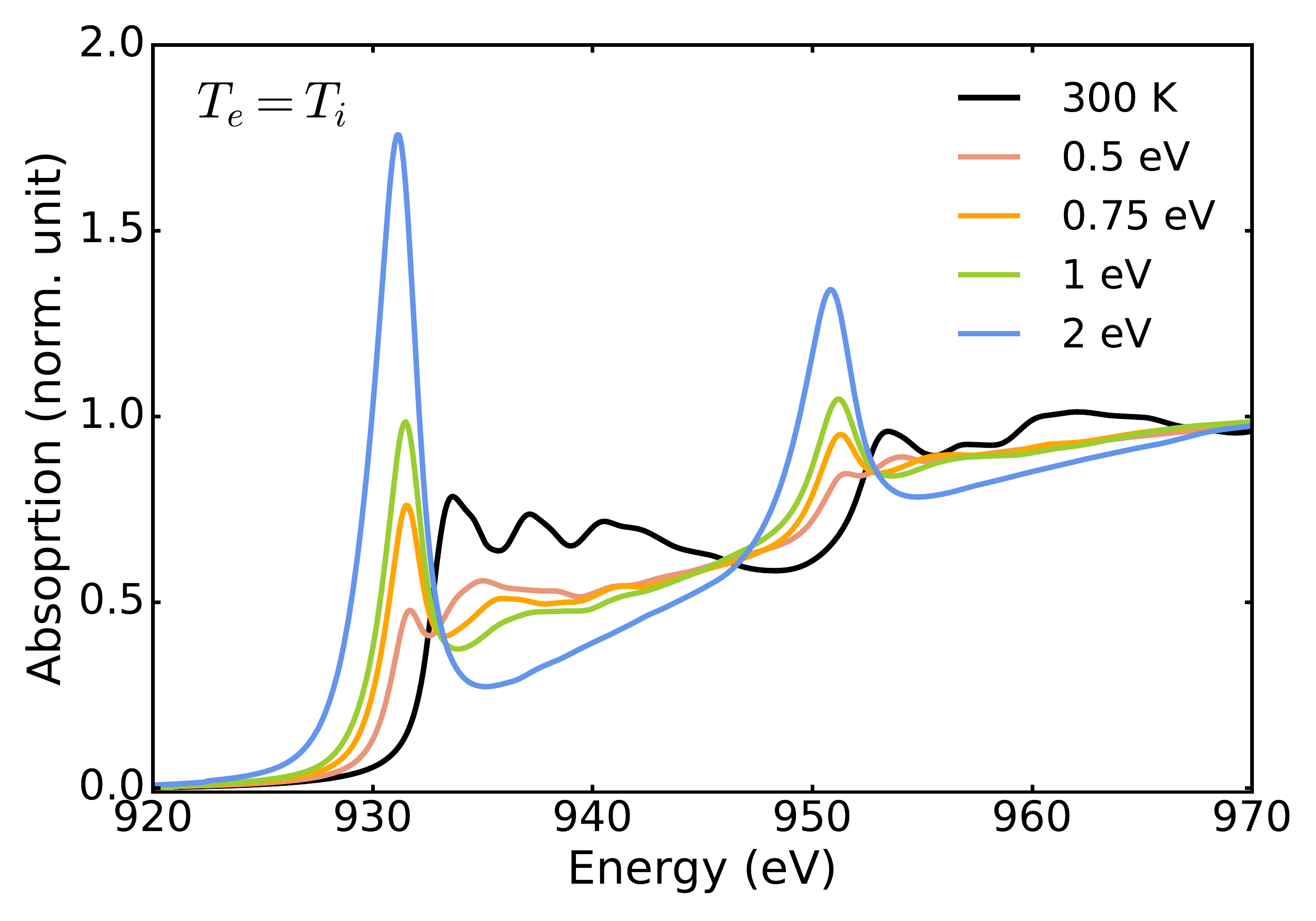}  
  \vspace*{0.5cm}\caption{(color online). XANES spectra at ambient conditions and for different
  temperatures in equilibrated situations ($T_e=T_i$) for the expanded density $\rho$ = 6 g/cm$^3$.}
  \label{XANES_6gcc}
\end{figure}

Figure \ref{XANES_6gcc} shows the computed XANES spectra for the expanded case at $\rho=6$
g/cm$^3$ and for equilibrated temperatures $T_e=T_i$ up to 2 eV. We also observe the rise of the
pre--edge structure with the electronic temperature $T_e$. Its amplitude value for a given temperature
is slightly higher compared to XANES calculations at the solid density $\rho_0$ reported in Figures
\ref{Cu_rho0_X_Te_Ti} and \ref{Cu_rho0_X_Te}. The pre--edge is also thiner and the absorption
level is lower above the absorption edge.

\begin{figure}
  \centering
  \includegraphics[width=\linewidth]{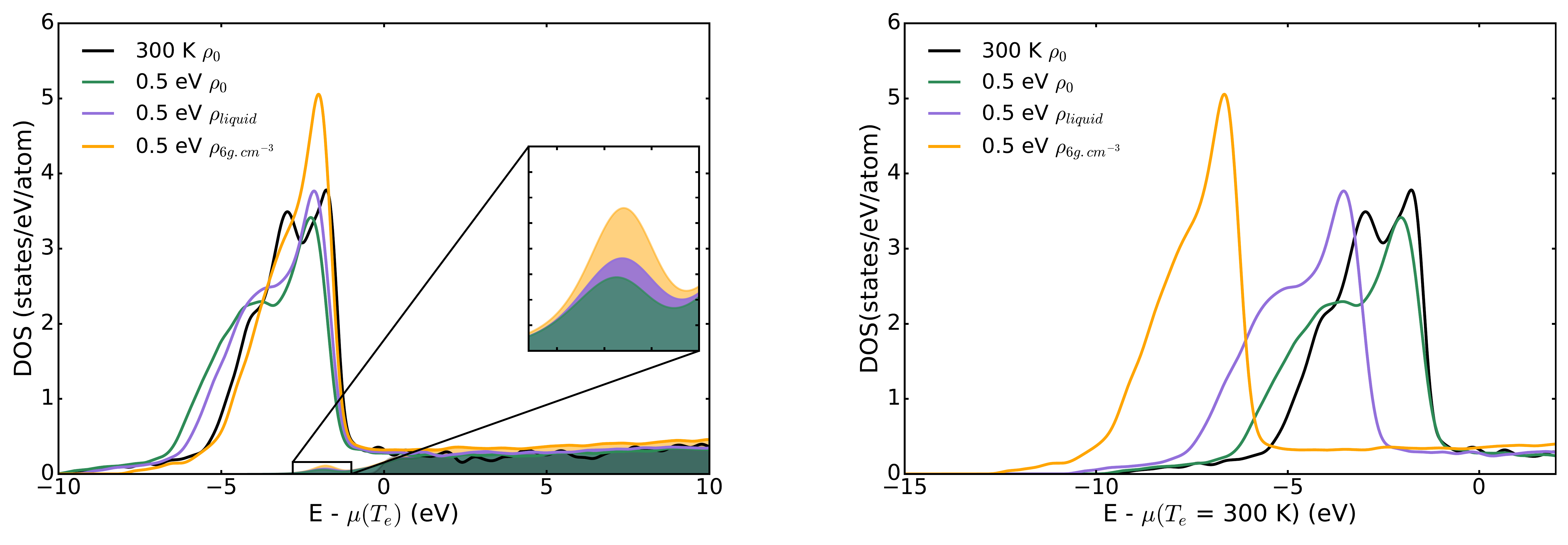}  
  \vspace*{0.5cm}\caption{(color online). left panel: Density of States (DOS) of copper at different densities
  for $T_e=T_i$ = 0.5 eV and compared to the one at ambient conditions. They are represented as a
  function of $\epsilon - \mu(T_e)$. The colored areas correspond to the vacant electronic states. 
  The inset shows a zoom of the region inside the square around -2.5 eV. Right panel : same DOS represent as a function of constant 
  $\epsilon - \mu(T_e=300K)$}
  \label{DOS_frho_mu}
\end{figure}

To interpret these observations, one can refer to the Figure \ref{DOS_frho_mu} showing the DOS for
several densities as a
function of $\epsilon - \mu(T_e)$. It appears that the $3d$-band is not strongly modified, with both
temperature and density. Therefore, the Fermi-Dirac distribution has similar impact on the $3d$-band
occupation. This is clearly visible in the shaded areas showing the unoccupied DOS. The
pre--edge structure keeps a similar shape in the absorption spectra at these lower densities (\textit{e.g.}
at $\rho=6$ g/cm$^3$ in Figure \ref{XANES_6gcc}).

Looking more closely at Figure \ref{DOS_frho_mu}, the $3d$-band gets thinner at lower density
and leads to a higher amplitude of the right side of this electronic band. Consequently, for a given
temperature, the pre--edge amplitude becomes higher when the density decreases.

\section{XANES spectra shifts}

The positions of the pre--edge and other peaks above the edge in the absorption spectrum are
connected to the energy difference between the DOS and the $2p$ orbital. Both undergo variations
with temperature and density that we have to take into account to describe a possible energy shift of
the absorption features when modifying the thermodynamical conditions. This is mandatory to be able
to compare theoretical and experimental spectra.

\subsection{DOS and chemical potential shifts with temperature and density}
\begin{figure}
  \centering
  \includegraphics[width=0.7\linewidth]{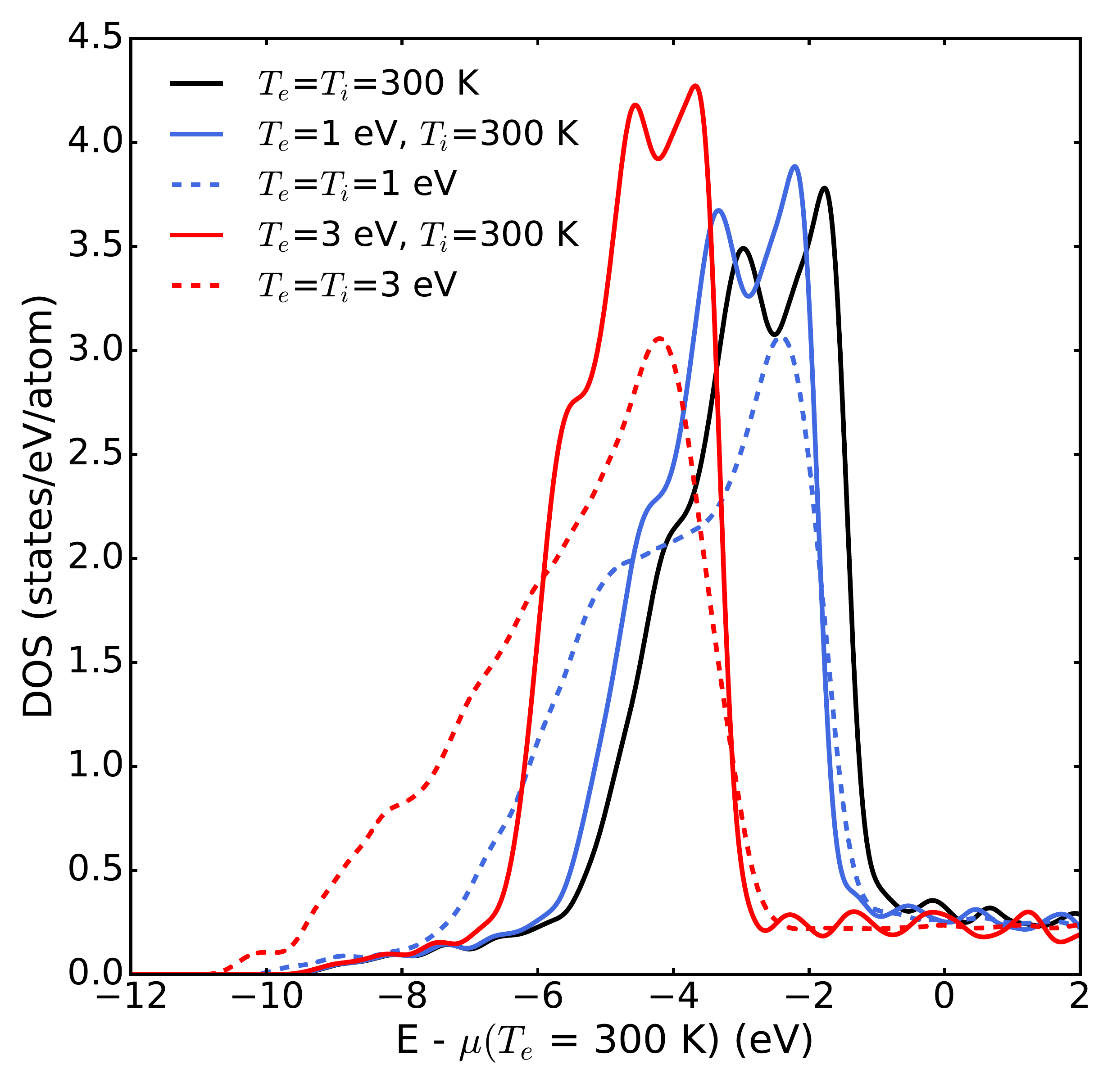}
  \vspace*{0.5cm}\caption{(color online). Density of States (DOS) of copper at solid density $\rho_0$ for
  both thermally equilibrated ($T_e=T_i$) and strong out-of-equilibrium ($T_e \gg T_i=$ 300 K) conditions
  up to 3 eV.}
  \label{Cu_rho0_eq_vs_HE}
\end{figure}

A negative shift of the DOS is observed when the density is fixed and the temperature is increased.
This effect is shown in the Figure \ref{Cu_rho0_eq_vs_HE} for the solid density $\rho_0$ and different
temperatures, in both thermally equilibrated ($T_e=T_i$) and strong out-of-equilibrium ($T_e \gg T_i$
= 300 K) situations. For a given electronic temperature, the energy shift is identical for these two situations. 

\begin{figure}
  \centering
  \includegraphics[width=0.7\linewidth]{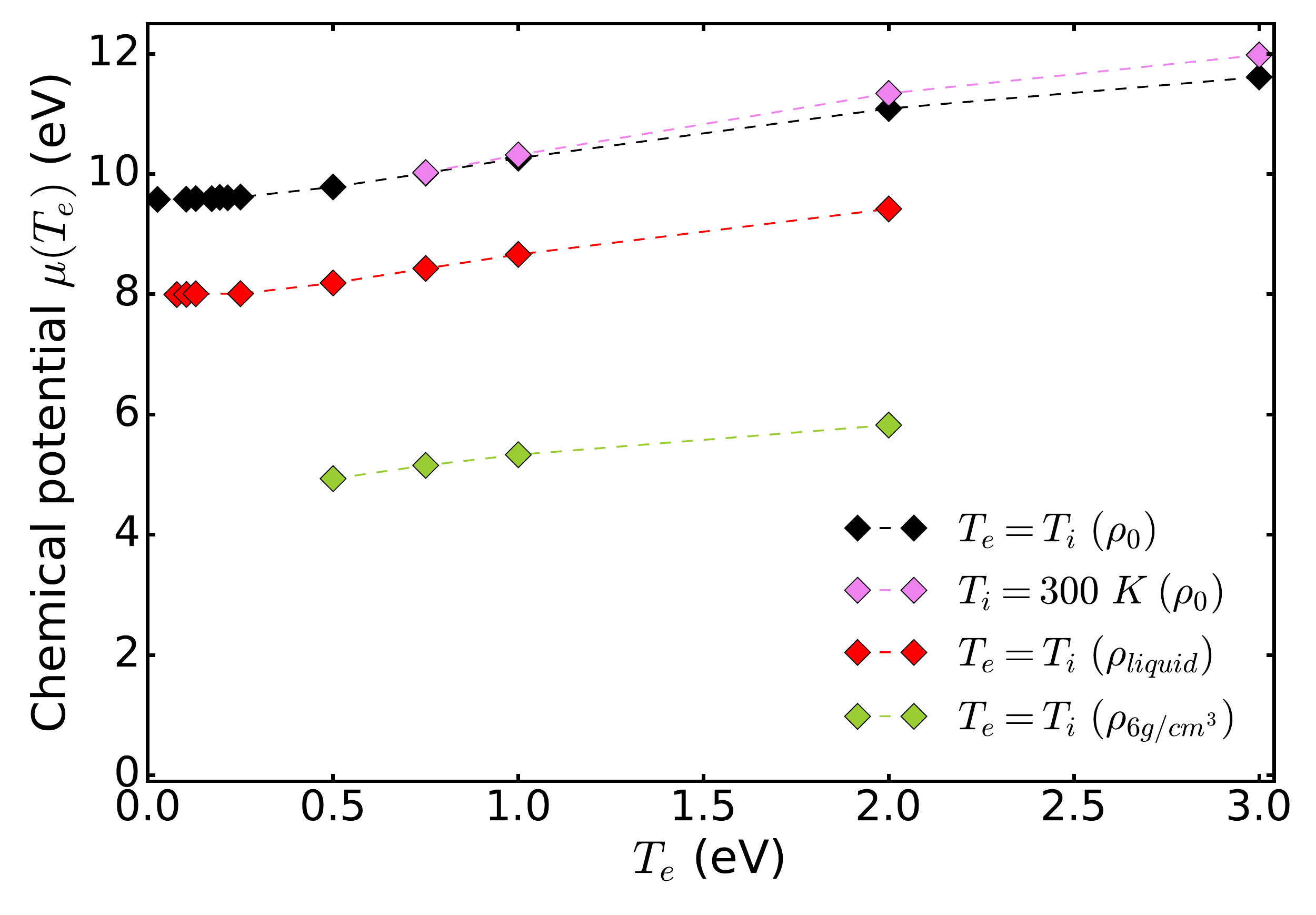}  
  \vspace*{0.5cm}\caption{(color online). Evolution of the chemical potential $\mu(T_e)$ with the temperature 
   of the electrons $T_e$ for three densities of copper $\rho = \rho_0, \rho = \rho_{liquid} =$ 8 g/cm$^3$ and
   $\rho = $ 6 g/cm$^3$.}\label{Cu_mu}
\end{figure}

The energy shift of the chemical potential $\mu(T_e)$ is directly correlated to the DOS one, since
$\mu(T_e)$ is obtained by the conservation of the total number of electrons  $N_e = \int g(\varepsilon,T)
f(\varepsilon,\mu(T_e),T)d \varepsilon\label{eq:partncons}$ with $g(\varepsilon,T)$ the electronic DOS
and $f(\varepsilon,\mu(T_e),T)$ the Fermi-Dirac distribution. As can be observed on figure \ref{Cu_mu},
the chemical potential increases with temperature and exceeds from around 2 eV its ambient value at $T_e=$ 3 eV 
and this is true for both cases $T_e=T_i$
and $T_e \gg T_i=$ 300 K. Given the similar motions between these two cases, we might reasonably assume that the 
number of electrons in these localized $3d$ bands is not different for a given temperature. 
The broadening of the Fermi-Dirac distribution and the negative DOS shift lead to an increase of the chemical potential.
This was observed for transition metals with localized $d$ states \cite{Holst_2014, Bevillon_2014}.
This is the opposite of the motion of the chemical potential computed for aluminum and this could
be traced back to the free-electrons like behaviour of aluminum \cite{Recoules_2009}.

When the density is lowered, the density of states is also shifted towards lower energies for a given temperature
as can be seen in the right panel of figure \ref{DOS_frho_mu}.

In parallel, the chemical potential drops due to the
conservation of the total number of electrons $N_{e}$ (see Figure \ref{Cu_mu}). In the expanded case, the $3d$ states
are even more localized than for the case at normal density. This was already observed for aluminum, where delocalized
states becomes more atomic like when density decreases. As a consequence, the chemical potential has to decrease to recover
the right number of electrons for a given temperature \cite{Levy_2012}. This is a density effect. For a given density, 
even for the lowest density explored here, the chemical
potential still increases with temperature. 

\subsection{\textbf{\textit{2p}} orbital relaxation}
When changes in temperature and/or density occur, the $2p$ orbitals relax to different energies. This
can result in possible XANES spectra shifts if the probed valence electronic states do not shift accordingly.
This was observed for expanded aluminum and molybdenum \cite{Levy_2012, Dorchies_PRB2015}.

\begin{figure}
  \centering
  \includegraphics[width=0.7\linewidth]{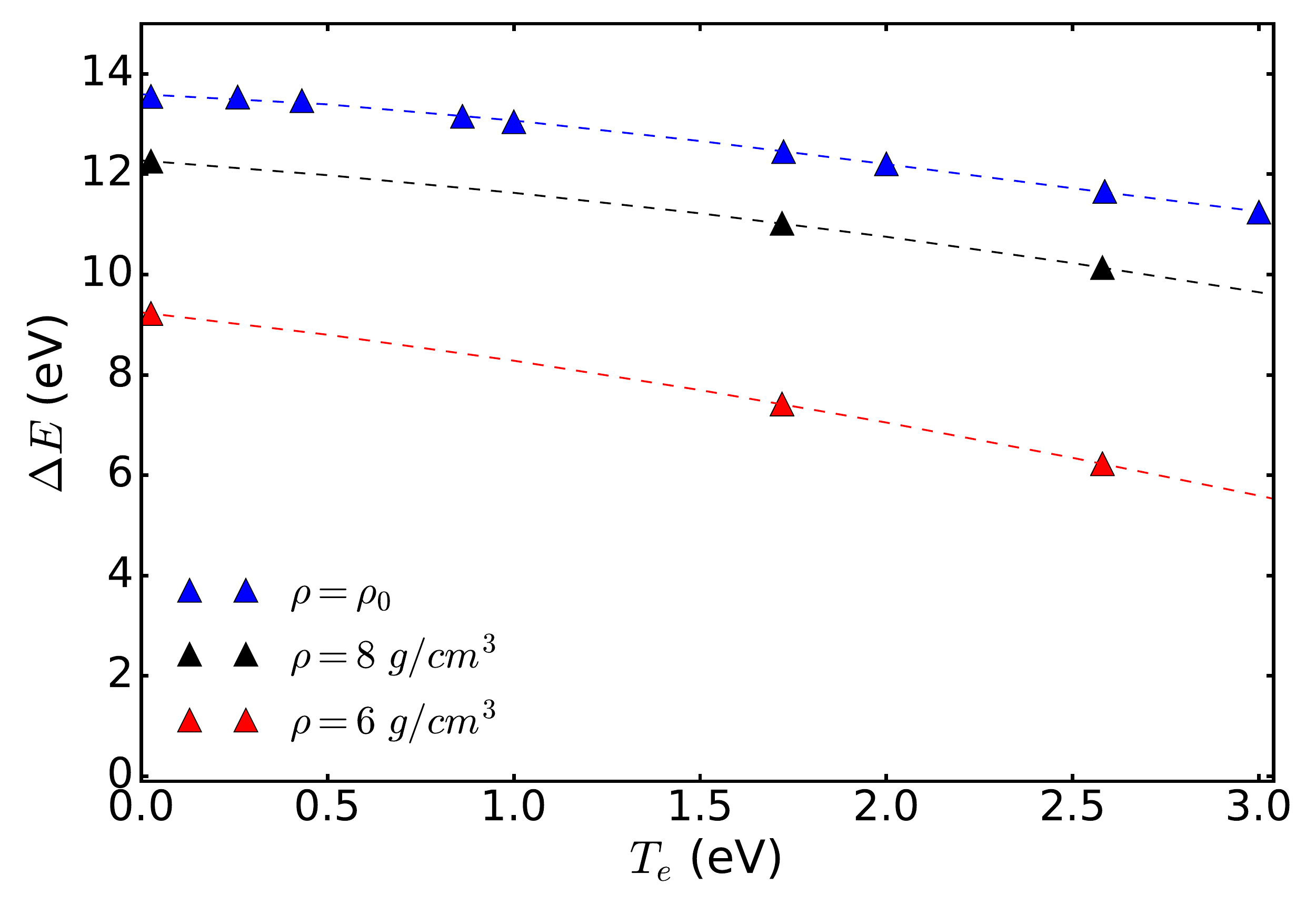}  
  \vspace*{0.5cm}\caption{(color online). Evolution of the energy difference $\Delta E$ between $2p$ orbital
  energies when considered in the frozen core or not. $\Delta E$ is defined as $E(2p_{relaxed}) - E(2p_{atomic})$
  where $E(2p_{atomic})$ corresponds to the value when the $2p$ orbital is frozen.}
  \label{2p_relax}
\end{figure}

We recall that all XANES calculations involve the frozen core approximation. The generation of the PAW
dataset using Atompaw and the energy value of the core orbitals result from an atomic calculation. To take
into account the relaxation we generate a different PAW data set where only $1s$ states are frozen. Relative
variations $\Delta E$ of relaxed $2p$ orbital energies compared to the atomic value are shown in
Figure \ref{2p_relax}. We see here that $\Delta E$ decreases when $T_e$ rises for each density considered,
meaning that $2p$ orbital gets closer to the nucleus. This can be understood by the reduction of the electronic
screening which leads to a more attractive electron-ion potential.

In these conditions, the highest shift of the $2p$ orbital is $-2.3$ eV for a temperature of 3 eV. This is almost
the same value as the DOS shift illustrated on Figure \ref{Cu_rho0_eq_vs_HE}, and this holds for all the
temperatures shown here. This is why we do not observe a visible shift on the XANES spectra presented
before in the Figures \ref{Cu_rho0_X_Te_Ti} and \ref{Cu_rho0_X_Te}.

Regarding the density effect, $2p$ orbital energy also decreases when the density is lowered. Both the DOS
(shown in Figure \ref{DOS_frho_mu}) and the $2p$ orbital move and almost in the same proportion (the maximum
difference is 0.2 eV between the DOS and the $2p$ shifts at $\rho$ = 6 g/cm$^3$), which results in an
unchanged position of the XANES spectra.

\section{Extracting $\bf{T_e}$ from x-ray absorption spectra}
In warm dense Be \cite{Schwanda_1992} and Al \cite{Dorchies_PRBb2015}, previous studies have
demonstrated that the temperature induced $K$ edge broadening could be fitted with a Fermi-Dirac
distribution function to retrieve the value of the electronic temperature $T_e$. This methodology applies
for relatively flat energy profile of the electron DOS. It cannot be used for copper whose DOS is dominated
by the strongly localized $3d$-band.

We have evidenced that both density and temperature have influence on the electronic occupations
dictated by $f(\epsilon,\mu(T_e))$. At the end, the variations in $\mu(T_e)$ and the resulting changes
in $f(\epsilon,\mu(T_e))$ always lead to the depopulation of electrons of the $3d$-band, giving a
pre--edge structure in the XANES spectra. The same phenomenon is observed for all the calculations
we have performed, $i.e.$ both ($T_e$,$T_i$)temperatures up to 3 eV, and densities ranging from solid density $\rho_0$
to densities as low as $\rho=$ 6 g/cm$^3$.
The question is then can we use this pre--edge to deduce the value of the electronic temperature?

For this purpose, we consider the value of the spectral integral of the pre--edge.
This integral is proportional to the number of unoccupied electronic states in the DOS. Figure
\ref{Cu_rho0_X_Te_Ti} shows through the hatched area the energy region considered for such pre--edge
integration (from 922.5 eV to 932.5 eV). The choice for integration region is arbitrary.
 The value of the integral of the cold spectrum in this spectral region
is subtracted to the integral in each case. The integral is not equal to zero in the cold case even if
no electron is thermally excited because of the spectral linewidth and the experimental broadening of the absorption edge.
Spectral resolution and natural linewidth could modify the height this is why we choose to use the integral instead.
Note that we compute the evolution of the pre-edge area refered to cold case. This, also, removed potential experimental resolution and natural linewidth effect assuming that there are the same for the different conditions explored here.
Looking at the comparison between experimental and theoretical spectra for cold and hot conditions presented in figure 4.b in our previous paper \cite{XANES_Cu}, one can see that the agreement is really good for both temperatures showing that these approximations seem reasonable.

\begin{figure}
  \centering
  \includegraphics[width=0.7\linewidth]{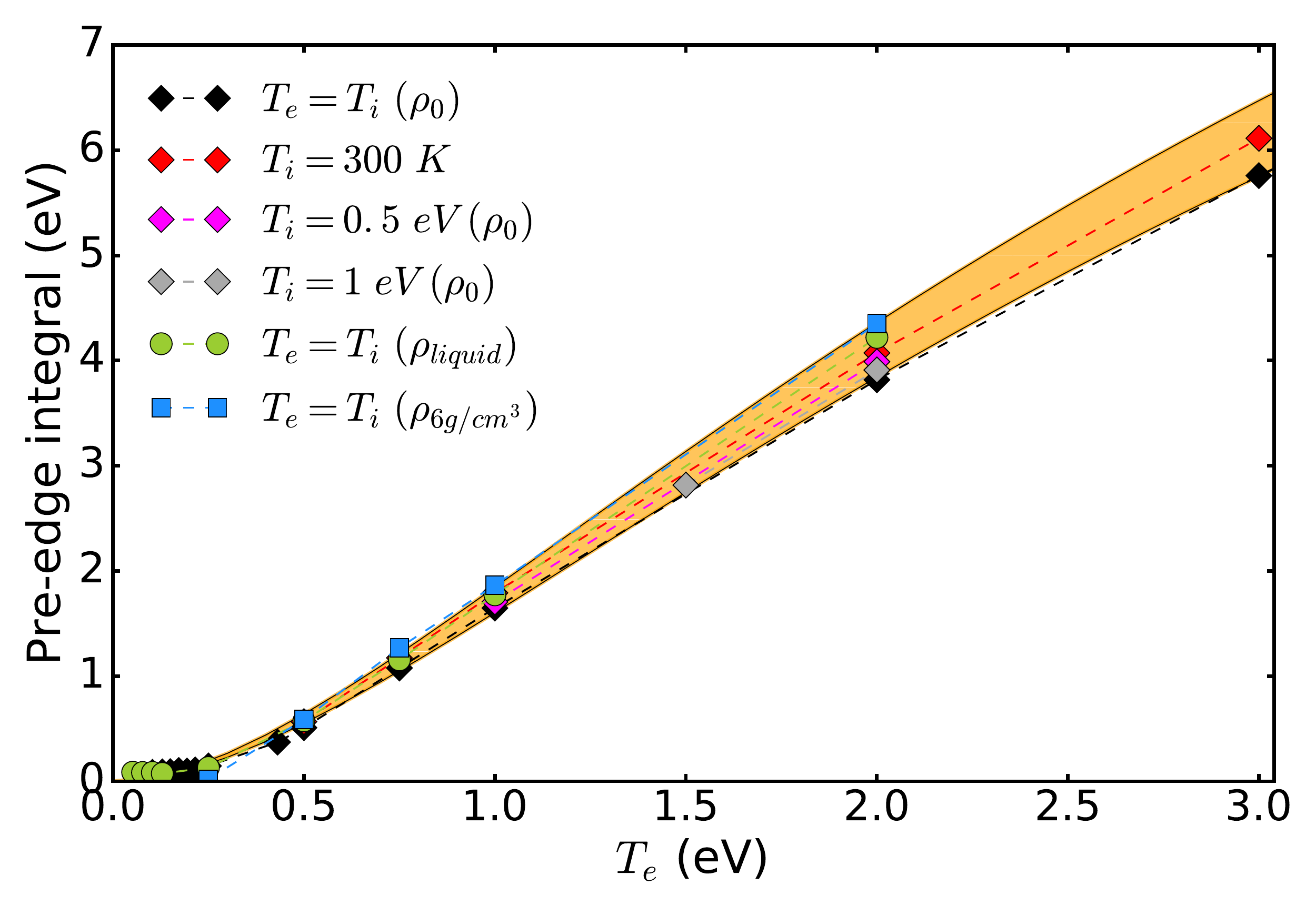}  
  \vspace*{0.5cm}\caption{(color online). Evolution of the pre--edge integral as a function of the electronic
  temperature $T_e$ for all the simulated conditions.}
  \label{pre-edge_vs_Te}
\end{figure}

Using our \textit{ab initio} XANES calculations performed for several temperatures at thermal equilibrium $T_e
= T_i$, out-of-equilibrium $T_e \neq T_i$ and for three different copper densities, we have computed the
value of the pre--edge integral for each spectrum and so for each electronic temperature $T_e$. The
results are presented in the Figure \ref{pre-edge_vs_Te}. Below $T_e = 0.2$ eV, the  pre--edge is too weak to be quantified and its evolution is not monotonous.Above 0.2 eV, the evolution of the pre--edge integral with $T_e$ is monotonous. 
More remarkable, it does not depend significantly on the equilibrium versus non equilibrium condition,
and depends little on the density. The amplitude of the pre-edge changes but not the area of the
pre-edge. At the end, the difference in the integral value reaches $\sim$ 15 $\%$ at 6 g/cm$^3$ compared to 8.9
g/cm$^3$. 

Above $T_e = 0.2$ eV, as the evolution of the pre--edge integral is monotonous, it is
possible to establish an unambiguous relationship between $T_e$ and the pre--edge integral. This relationship has been recently
used to measure the electron temperature dynamics of laser heated copper with a picosecond time
resolution \cite{XANES_Cu}. To be able to recover directly electronic temperature from experimental spectra, 
the requirement is to use for the integration exactly the same region as was chosen for theoretical spectra.

\section{Conclusion}
In summary, we calculated X-ray absorption spectra near $L_{2,3}$ edges of warm dense copper based
on first principle electronic structure calculations and molecular dynamics simulations. By introducing a
variable width to model the linewidth, we obtain a good description of the experimental
XANES spectrum and its modulations at room temperature. 

We explain the modifications induced by both electron and ion temperatures and density on XANES spectra at
the microscopic level. The shape is modified in two ways.  First, as copper is a noble metal
with almost filled $3d$ states at ambient conditions, they do not appear on XANES spectrum near $L$ edge
at ambient condition. But, when the temperature of the electrons $T_e$ increases, the $3d$ states are
partially depopulated, leading to a pre--edge on the absorption spectrum and a reduction of the absorption behind the edge.
The amplitude and shape
of the pre--edge is directly linked to the modification of $3d$ electronic population. 
Second, the modulations above the edge disappear when the temperature increases as solid copper turns to a liquid state. 
This opens a promising perspective to the experimental study of phase transition dynamics in non-equilibrium warm dense transition metals. 
This observation would require further investigation. In the range of
investigation (temperature up to 3 eV and density as low as 6 g/cm$^3$), we do not observe any energy
shift of the XANES spectra, as the $2p$ core level shift is balanced by the DOS one. 

The spectral integral of the pre--edge can be used to deduce the electronic temperature $T_e$ for a wide
range of temperatures and densities. This leads to a useful tool to associate an electronic temperature $T_e$ to
experimental XANES spectra. This tool can be used for any measurement of XANES spectra for femtosecond laser heated copper
 without the need of additional calculations keeping in mind that the same integral region has to be chosen to analyze the experimental spectra. The same considerations can be applied regarding the $d$-band of the other
noble metals (Ag and Au). It could be possible to extract the electronic temperature in a similar way for other transition
metals with localized $d$ states.

We believe that for any material, there should be a direct connection between the electronic temperature and the shape on the XANES spectra. Then, there is always a way to directly \textit{read} the temperature on XANES spectra. The signature will depends on the DOS. As soon as there is no localized states near the Fermi level, the temperature could be extracted directly by fitting a Fermi-Dirac function to the edge.
If there is localized states near the Fermi level as for the transitions metals, the temperature will appear as a pre--edge or a shoulder. For all cases, one need first dedicated simulations to quantify the change induced by temperature.

\begin{acknowledgments} 
 The authors are extremely grateful to M. Torrent for his precious help 
for the implementation in the \textsc{Abinit} code.
\end{acknowledgments}

%
\end{document}